# Is the Juice Worth the Squeeze? Machine Learning in and for Agent-Based Modelling

A Preprint

**Johannes Dahlke**
Innovation Management
University of Hohenheim
Stuttgart, Germany
johannes.dahlke@uni-hohenheim.de

**Kristina Bogner**
Innovation Economics
University of Hohenheim
Stuttgart, Germany
kristina.bogner@uni-hohenheim.de

**Matthias Müller**
Innovation Economics
University of Hohenheim
Stuttgart, Germany
m_mueller@uni-hohenheim.de

**Thomas Berger**
Land Use Economics
University of Hohenheim
Stuttgart, Germany
i490d@uni-hohenheim.de

**Andreas Pyka**
Innovation Economics
University of Hohenheim
Stuttgart, Germany
a.pyka@uni-hohenheim.de

**Bernd Ebersberger**
Innovation Management
University of Hohenheim
Stuttgart, Germany
ebersberger@uni-hohenheim.de

*March 20, 2020*

## Abstract

In recent years, many scholars praised the seemingly endless possibilities of using machine learning (ML) techniques in and for agent-based simulation models (ABM). To get a more comprehensive understanding of these possibilities, we conduct a systematic literature review (SLR) and classify the literature on the application of ML in and for ABM according to a theoretically derived classification scheme. We do so to investigate how exactly machine learning has been utilized in and for agent-based models so far and to critically discuss the combination of these two promising methods. We find that, indeed, there is a broad range of possible applications of ML to support and complement ABMs in many different ways, already applied in many different disciplines. We see that, so far, ML is mainly used in ABM for two broad cases: First, the modelling of adaptive agents equipped with experience learning and, second, the analysis of outcomes produced by a given ABM. While these are the most frequent, there also exist a variety of many more interesting applications. This being the case, researchers should dive deeper into the analysis of when and how which kinds of ML techniques can support ABM, e.g. by conducting a more in-depth analysis and comparison of different use cases. Nonetheless, as the application of ML in and for ABM comes at certain costs, researchers should not use ML for ABMs just for the sake of doing it.

**Keywords** Agent-based Modelling · Multi-Agent System · Machine Learning · Reinforcement Learning · Output Analysis · Systematic Literature Review · Research Methods

## 1 Introduction

The interest in and the application of the method of agent-based modelling (ABM) has increased tremendously in recent years. Despite the fact that the underlying nature of ABMs (e.g. concerning their aims and goals, their characteristics, or their relationships with other simulation techniques) still is not generally agreed upon, this method is increasingly applied by different researchers from various disciplines in many diverse fields of application. And this even well beyond the usual ones (North & Macal, 2007). As Macal nicely puts it, the general consensus seems to be: "We do not know what agent-based modeling is, ... but we know we need it!" (Macal, 2016, p. 145). What we do know, however, is that ABM allows to model and analyze complex adaptive systems consisting of autonomous, decision-making agents that interact with each other and their environment (Pyka & Fagiolo, 2007). The unique feature, especially in contrast to other simulation methods, is the agent perspective and the inherent bottom-up approach (Axelrod, 1997). ABM helps to understand both agents' and aggregate behavior and explain how these behaviors lead to large-scale outcomes and emergent phenomena (North & Macal, 2007). With advanced computational possibilities and an increasing interest in larger, more and more data driven models, agent-based models tend to become increasingly complex to build, analyse and understand (Macal, 2016). In such models, problems are often related to the increase in used and produced (big) data, especially concerning (1) the empirical parameterization of the model, (2) the most efficient construction of the mathematical computer model, (3) model validation and verification, as well as (4) the sensitivity and output analysis.

Despite all the joy about the apparently bright future of ML in and for ABM, three things must not be forgotten: First, current contributions are often conceptual and theoretical in nature or only operating in a very narrow field, lacking a comprehensive synthesis / overview of what's actually been done in current applications of ABM or what might be doable and interesting in the future. Second, we think that the proposed claims need to be considered with caution and potential downsides of the use of ML in ABM need to be discussed properly. Finally, while the previously mentioned studies exhibit great expertise and advance the discourse by presenting bold visions of ML and ABM, individual exploratory studies of applied ML in ABM only present a limited picture of potential uses. We need to rely on the collective knowledge base among authors involved in this methodological field. Thus, we seek to analyze how researchers from different (and distant) academic fields actually use the combination of methods and for what purpose they do so. To our knowledge, such a synthesis has not been put forward so far. To generate this comprehensive overview, we conduct a systematic literature review (SLR) and, based on a wide range of keywords, sample 397 journal publications working at the interface of ML and ABM. We theoretically derive a clustering scheme based on common process definitions of ABM, current issues faced in ABM research and a classification of machine learning algorithms and tasks. By subsequently clustering the literature among these dimensions, a picture of the currently most prominent applications and use cases emerges. Through this process, we develop a comprehensive understanding of the gains and pains of using ML in and for ABM, which helps us in properly assessing the claims found in the literature, and finally allowing us to answer: 'When is the juice worth the squeeze?'. To reach this end, we set out to investigate the following two research questions:

**RQ1: How has machine learning been utilized in and for agent-based models so far?**

**RQ2: What are the potential (dis-)advantages of using ML in and for ABM?**

Our paper is structured as follows: Section 2 sets the scene by giving a brief overview of both the method (and issues) of agent-based modelling as well as some of the most important machine learning techniques. Section 3 presents how the structured literature review is conducted in this paper and according to which characteristics we structured the literature and the results. Subsequently, section 4 presents and critically discusses the results of our analysis, by providing a presentation of how ML has been used so far, a presentation of some cases of more and less promising fields of application of ML in and for ABM, as well as a critical reflection on claims from the literature. The concluding Section 5 summarizes our article and proposes some avenues for further research.

## 2 Theoretical Background: On Agent-Based Modelling and Machine Learning

In this section, we describe the theoretical foundations of agent-based modelling (2.1 and 2.2) and machine learning (2.3) in order to derive the classification scheme used for categorizing the literature found in the SLR.

### 2.1 On Building and Running an Agent-Based Model - ABM steps

Within the broad field of simulation techniques, the so-called agent-based modelling approach has gained increasing momentum over the last decades, not only for economics but also in many other scientific disciplines. In this context, "ABMS represents a fundamentally new simulation and modelling technique that offers the potential to solve problems that are not robustly addressed by other methods." (Macal, 2016, p. 153). The ABM approach takes the perspective of the system building elements and focuses on the actions and interactions of these entities. As a consequence, ABM offers a natural description of real-world systems through which we can incorporate the dynamics of emerging phenomena within a powerful and flexible simulation environment (Bonabeau, 2002). Up until now, a comprehensive and precise common definition of agent-based modeling cannot be found in the literature and is quite unlikely to emerge in the future (Macal, 2016). What is actually defined as an agent-based model (or even only as an agent) can vary widely, being highly dependent on, e.g., the background of the modeler. As many different, heterogeneous groups from various disciplines (such as management science, economics, social science, geography, ecology, behavioral and organizational sciences, as well as complexity science) are engaged in agent-based modelling, quite different definitions, understandings and research traditions emerged over time (Macal, 2016). In general, an agent-based model is a model "composed of individual entities that have autonomous behaviours and are different from one another, having diverse characteristics and behaviours over a population" (Macal, 2016, p. 149). Such a model typically consist of a set of autonomous agents, equipped with a set of attributes and a set of specified behaviors, connected to other agents and their environments by certain relationships (Pyka & Fagiolo, 2007; Taylor, 2014). These agents, their environment and their relationships therefore compose a system with clearly defined boundaries, inputs and outputs (Taylor, 2014). Despite this variety in and of definitions, there exists a general consensus beyond the boundaries of disciplines concerning the most important steps generally conducted when creating, implementing and analyzing agent-based models and their behavior (Gilbert & Troitzsch, 2005). The general steps typically being performed are the following four: During 1) the conceptualization and empirical parametrization, we start with a simple definition of the simulation goals. Based on this, we formulate a thought model for which we already identify and determine the key assumptions (for example based on empirical data or the literature). This is followed by 2) the construction of a mathematical computational model, in which we implement our thought model in a computer program. This is usually done through specialized simulation environments such as for example NetLogo (Wilensky, 1999), LSD - Laboratory for Simulation Development (Valente, 2008), Repast – Recursive Porous Agent Simulation Toolkit etc. Next, within 3) the validation and verification phase, the modeler needs to ensure that the thought model has been implemented correctly and that the models output and behavior corresponds to the intended or foreseen behavior of the thought model. Already small mistakes during the implementation of the model in a simulation program can end up to unintended results which will distort the simulation results and in the worst case can lead to false conclusions (Ormerod & Rosewell, 2009). Finally, during 4) the output analysis, the modeler analyzes model outcomes and causal relationships, e.g., by looking for emerging patterns for specific combinations of relevant parameters to derive implications or a narrative from output data.

### 2.2 On Building and Running an Agent-Based Model – Problems in and Challenges for ABMs

Especially due to the facts that ABM is a relatively new and promising method applicable in many different fields in a myriad of ways, it comes as no surprise that this ever-evolving method faces several research challenges. These challenges necessarily have to be addressed for allowing to unfold ABMs full potential. "(...) (I)ndeed ABMS is a revolution in progress, although more of a promise of what ABMS could accomplish than a realized potential" (Macal, 2016, p. 146, referring to Bankes, 2002). As ML is most of the time used for supporting or improving ABMs, in the following, we're going to present three major challenges ABM is currently facing (Macal, 2016)) and use these for deriving the overall goals of applying ML in and for ABM we later use in the classification scheme.

First, one of the major interests in ABMs is a more accurate representation of (heterogeneous) human behavior and the aggregate effects of these behaviors. While ABMs theoretically offer the possibility of simulating and analyzing more realistic and heterogeneous agents, than, e.g. mainstream neo-classical models, many ABMs still show a rather homogeneous and static set of agents, which are often modeled rather unrealistic in the first place and also limited in their adaptability, i.e. their ability to learn and change behaviors during the simulation. Therefore, one of the major challenges to not only increase ABMs explanatory power, but also their credibility, is to develop better representations of agent behavior, i.e. to make agents' behavior more realistic (Macal, 2016). This *behavioral modelling challenge* can be addressed by (i) incorporating more realistic assumptions on agent characteristics and behavior based on empirical data by inferring agent behaviors from data streams, e.g., real-time data extraction from social media. It can also be addressed by (ii) increasing agents' adaptability and their ability to learn during the simulation, e.g., by explicitly modelling learning agents. And (iii) behavior models can be modelled more realistic by calibrating and validating them against real-world data (Macal, 2016). Addressing the behavioral modelling challenges would therefore serve the overall goal to improve the models' accuracy and the (quality of) assumptions built into the model.

Two other related major challenges are the *large-scale agent-based modelling challenge* and the *simulation analytics challenge*. Handling the model, understanding the exact model behavior and extracting meaningful insights from simulation results is a quest all modelers have to meet. Enabled by the constant improvement of computational possibilities within the last decades, researchers aimed at increasing the explanatory power of ABMs by using a more accurate, data-driven representation of the real world. ABMs have become larger and more complex. While these large-scale ABMs without doubt offer potentially fruitful insights, the complexity of models created through an ABM approach is limited by our capability to handle, understand and analyze the respective model and its output. If the size and the complexity of the model is reaching a level where we are no longer able to understand the processes involved, we cannot understand these artificial complex systems any better than we understand the real ones (Axtell & Epstein, 1994; Gilbert & Terna, 2000). Understanding and making use of the output created by such simulations is a challenging task. In this context, the large-scale agent-based modelling challenge implies both a *computing challenge* and a *research challenge*. The computing challenge relates to the problems of computational tractability and the computer's ability to handle the computational load of such large-scale models. The research challenge relates to the problem of the researcher to actually "dynamically balance simulation workloads, interact with running simulations, and efficiently collect model outputs for further analysis" (Macal, 2016, p. 153). Addressing the large-scale agent-based modelling challenge would therefore serve the overall goal of finding ways to cope with such large-scale ABMs, e.g., by finding ways of improving computational efficiency, for instance by reducing computational time or reducing computational load.

Albeit being closely related to the large-scale agent-based modelling challenge, the simulation analytics challenge rather deals with problems related to output analysis of the model (instead of problems of computational tractability and manageability). First of all, especially in more complex models with many different parameters, output data has to be stored and prepared in a way that it can be efficiently accessed by a human being or by a computer. What is more, key relationships have to be extracted and key parameters to be identified to understand major relationships. This understanding is a prerequisite for cleverly designing computational experiments to ensure computational tractability (Macal, 2016). Addressing the simulation analytics challenge would therefore serve the overall goal of enhancing computational tractability, improving the modelers understanding of model behavior and enabling the analysis of outcome data to find important parameters, clusters, causalities, patterns and potential anomalies.

Many researchers argue that these challenges ABMs are facing can quite naturally be addressed by use of machine learning techniques (Hoog, 2019; Edali & Yücel, 2019; Lamperti, Roventini, & Sani, 2018). In this context it can be argued that making use of ML in and for ABM can not only solve the presented problems but – what is more – help to realize ABMs enormous potential. Therefore, applying ML in and for ABM most of the time serves (at least) one of the goals presented in Table 1.

### 2.3 A Brief Overview of Machine Learning Algorithms (MLA)

Discourses around the topic of artificial intelligence (AI) are full of terminological confusions and loosely used buzzwords. Thus, we feel the need to preface this section with some definitory groundwork: Machine learning (ML) is a sub-category of artificial intelligence and describes algorithms that are able to learn from data without being equipped with specific programming of rules or models to follow (Domingos, 2012; Marsland, 2015; Pyle & Jose, 2015). These artificial computer systems are able to learn from parsing training data in order to deduce patterns and make inferences on unknown data. This ability to generalize beyond a set of training data is the key advantage of ML. For the purpose of this paper, we follow a (sometimes too) simplistic but widely acknowledged classification of ML algorithms and differentiate between supervised, unsupervised, semi-supervised and reinforcement learning (Marsland, 2015).

Generally, we can distinguish between machine learning algorithms (MLA) solving different types of tasks by asking whether or not the underlying database is labeled or unlabeled in nature. That is, whether or not observations of specific input data are connected to observations in selected outcome data *a priori*. Supervised MLA learn by parsing labeled data mainly for solving two major tasks: classification (assigning a categorical value depending on given input data) and regression for, e.g., predicting a continuous numerical value depending on given input data (Marsland, 2015). Classification techniques include decision trees (specific algorithms include ID3, C4.5, C.5) and random forests, naive bayes, bayesian belief networks and support vector machines (Friedman, Geiger, & Goldszmidt, 1997; Gupta, Gupta, & Singh, 2015). Both tasks may also be performed by different types of deep learning neural networks (NN) such as convolutional NN (CNN), recurrent NN (RNN), multilayer perceptron models (MLP) or recursive neural tensor networks (RNTN) (Battula & Prasad, 2013; Neelamegam & Ramaraj, 2013). Unsupervised MLA learn from unlabeled data and are mainly used for three objectives: clustering (assigning segments based on similarity of data points), anomaly detection (finding outliers) and association rules (finding if-then relationships between data points). A typical clustering algorithm constitutes k-means clustering, anomaly detection may be performed by using a low pass filter and association rules can be identified with so-called Apriori algorithms. All three tasks may also be performed using deep neural networks in the form of self-organizing maps (SOM) (Kohonen, 1997) or auto-encoders such as restricted Boltzmann machines (RBM) (Salakhutdinov, Mnih, & Hinton, 2007; Smolensky, 1986).

The techniques of supervised and unsupervised techniques may also be bridged by semi-supervised methods, including a controlling instance in the mix to label some data points initially before two opposing algorithms work on generating and discriminating more data for training purposes - as is the case with the general adversarial neural network (GAN) (Yang, Murata, & Amari, 1998). However, not all MLA can be categorized by the degree of supervision. A fourth class of techniques can be formed around the concept of reinforcement learning (RL). The unique proposition of this type of algorithm is the goal-oriented learning through iterative interactions with the environment (Sutton & Barto, 1988). Unlike in supervised procedures, the algorithm does not get supplied with the correct way through a label (i.e. target data) but only receives numeric rewards or scores, which judge the taken actions as good or bad based on the outputs. These rewards are to be maximized by the algorithm autonomously in an iterative trial-and- error approach not unlike the Markov decision process (Marsland, 2015; Parr & Russell, 1998). Because of this, RL is especially useful in situations where the algorithm cannot be supplied with a dataset of correct answers or when it is not meant to learn from prescribed (human) best practices. Contrary to the popular but partly misconceiving understanding of completely autonomous algorithms, conducting machine learning processes requires legwork to be done. Assuming a typical supervised machine learning process, the workflow starts by collecting and preprocessing data, which, as the research community is well aware of, may require specialist expertise as well as it can prove to be a time-consuming task to clean the data of noise and errors. Subsequently, the data is partitioned into a set of training data, a set for validation and a sample of target data (test data) on which the refined algorithm will finally be applied to. Depending on the size of the data at hand, the data is partitioned into chunks of 50:25:25 or 60:20:20 (Marsland, 2015). Next, it needs to be determined which characteristics of the collected data lend themselves best to be selected as inputs from the data in accordance with the computational task and model design at hand. An appropriate ML algorithm needs to be chosen as for any given computational problem (such as a classification task) since there are a number of algorithms performing differently depending on the context of application. This is why, in our classification scheme, we also sort the use-cases according to the computational problem at hand / the respective ML task (i.e. regression, classification, time-series forecasting, clustering, anomaly detection, association rules, dimensionality reduction and optimisation). Moreover, parameters for algorithms need to be set and continually refined in the training procedure of applying the ML solution to the validation datasets. Again, this step requires time and computational capacities. Before the final employment of the algorithm, the results of the algorithms need to be evaluated and validated through a comparison to other (real world) data samples. This, again, may require an expert in the loop (Marsland, 2015).

When working with ML algorithms, the size of the dataset is a crucial consideration. There is the question of how much data is needed in order to train and validate an algorithm. Of course, the most reasonable answer is that it depends. Namely, on the complexity of the task and the algorithm at hand. However so, it has been shown that more data can lead even weaker algorithms to outperform stronger algorithms with less data to train on. Thus, it may not primarily be the algorithm itself but the availability of data that drives performance in ML applications (Domingos, 2012). Large amounts of data are especially useful to mitigate the overfitting problem of ML algorithms. This problem arises whenever an algorithm becomes very effective in predicting outcomes based on the data it was trained on but fails to perform as well on new data it is given. The reason being that the algorithm has become biased towards structures in the training data, which are not representative of other samples (Domingos, 2012). It is possible to avoid this problem through the process of cross-validation. As more data likely covers more variations of data, the algorithm's ability to generalize predictive patterns for new data is enhanced. This is why, in the literature, we also find a range of applications in which ABMs are used for ML (and not the other way around). In these cases, researchers, e.g., use ABMs to produce data on which the ML algorithm can be trained on (Leottau, Solar,

& Babuška, 2018; Zhang, Olsen, & Block, 2007). When weighing classical statistics against ML, it quickly becomes clear that ML is better suited for wide data – that is, rich and unstructured data with a high number of input variables. In wide datasets, it is especially challenging to design a comprehensive statistical model to depict realistic influences. At this point, ML starts to excel classical statistics as it doesn't require to make any theoretical assumptions about the data at hand (Bzdok, Altman, & Krzywinski, 2018).

## 3 Methodology

In order to answer our first research question, we seek to investigate the status quo of the use of machine learning in ABM by following the procedure by Denyer and Tranfield (2009) and carry out a systematic literature review (SLR) based on a comprehensive sample of literature.

**Article Retrieval**

We define a number of keywords driving the systematic search process. Naturally, we draw from the two respective fields of literature at heart of our research, namely machine learning and agent-based modelling. First, we include the broader terms of machine learning, machine intelligence and machine translation. Again, we follow Marsland (2015) in distinguishing between the four learning paradigms of supervised, semi-supervised, unsupervised and reinforcement learning. For four paradigms we identify specific algorithms to include as keywords: For the supervised paradigm we include decision trees (Quinlan, 1986), random forests (Breiman, 2001), naive Bayes (Friedman, Geiger, & Goldszmidt, 1997; Kohavi & John, 1997), k-nearest neighbor (kNN) (Cover & Hart, 1967) and support vector machines (Burges, 1998; Wu et al., 2008). For the unsupervised paradigm, we include k-means (Jain, 2010) for clustering purposes, low-pass filter for anomaly detection and apriori algorithms to represent association rules (Wu et al., 2008). We include reinforcement learning (Sutton & Barto, 1988) as well as the most commonly used RL algorithm type of Q-learning (Watkins & Dayan, 1992) and the specific technique of gradient descent (Baird & Moore, 1999). Certain types of neural networks (NN) overarch the above-mentioned paradigms. Thus, we hope to catch the most common variations of NN by following the comprehensive overview of Baird and Moore (1999) to include the generic terms (artificial) neural networks (Schmidhuber, 2015) as well as deep learning (Hinton, Osindero, & Teh, 2006; Salakhutdinov, Mnih, & Hinton, 2007) but also the more specific forms of backpropagating NN (BPN), convolutional NN (CNN), feedforward NN (FNN), recurrent NN (RNN), deep belief networks (DBN), self-organizing maps (SOM) and – to include an auto-encoding algorithm – restricted Boltzmann machines (RBM) (Schmidhuber, 2015). In addition, we include generative adversarial networks (GAN) as an interplay of generating and discriminating NN (Goodfellow et al., 2014). The scientific method of agent-based modelling is characterized by terminological diversity. To represent this diversity of notions and emphasis, we include the following keywords to retrieve relevant literature: Beside agent-based modelling (Pyka & Fagiolo, 2007), we search for agent-based simulation (Polhill, Gotts, & Law, 2001), multi-agent simulation (Ferber, 1999; Gilbert & Troitzsch, 2005), multi-agent-based simulation (Edmonds, 2001), agent-based social simulation (Doran, 2001; Downing, Moss, & Pahl-Wostl, 2000), individual-based (configuration) modelling (Judson, 1994) and multi-agent systems (Bousquet & Le Page, 2004). While these terms may represent semantic differences between subject areas of research, Uhrmacher and Weyns (2009) argue that there is no clear methodological line of differentiation between the terms of ABM and multi-agent systems (MAS). Regardless of a missing consent concerning the appropriate notion for this scientific method, the understanding of ABM between the different roots is very similar – see also (Hare & Deadman, 2004) on this issue. We use the bibliographic search engine Scopus to conduct our search. The search query includes all of the aforementioned terms in a lemmatized form (indicated by an asterisk, for example, instead of 'neural network', we search for 'neural net*'). In order for a publication to show up in our sample, it needs to exhibit at least one term related to the area of ML (connected with Boolean 'OR'), while it simultaneously (connected with Boolean 'AND') needs to represent at least one term related to the method of ABM (connected through Boolean 'OR'). We search for these terms within the titles, keywords and abstracts of articles published in peer- reviewed journals. We limit our search further by only considering publications written in English language. We purposely choose a two-pronged approach with two distant fields in (managerial) economics and environmental science in order to gain a broader spectrum of potential applications of ML in ABM that are context dependent. Moreover, the selected subject areas in line with the expertise of the authors. We include only publications associated with the subject areas of social science (SOCI), business, management and accounting (BUSI), economics, econometrics and finance (ECON), environmental science (ENVI), energy (ENER) and agricultural and biological science (AGRI). The full search query used is presented in Appendix A.

**Article Selection / Final Sample**

We follow Moher et al. (2009) and represent the selection process for publications in our final sample by the schema of the 'Preferred Reporting Items for Systematic Reviews and Meta-Analyses' (PRISMA). In a first step, relevant articles are extracted through Scopus based on our search query. We included articles published until the end of 2019, this yields 397 articles. Next, the sample is split among co-authors to evaluate the abstracts of publications for relevancy. This procedure is double-checked using a four-eye principle. We identify 190 relevant articles to be included in the in-depth analysis. 207 articles have been excluded due to irrelevancy, e.g., if the authors used ABM in and for ML and not the other way around. For potentially relevant publications that don't allow to be assessed judging only by the abstract, we perform a check for eligibility by retrieving and analyzing the full paper.

**Analysis of Final Sample**

In order to coherently cluster our final sample, we devise a classification scheme, which is closely linked to the theoretical considerations in Chapter 2. We end up with four dimensions: We distinguish between the different steps of the ABM process (presented in 2.1), the goals of applying ML derived from the general challenges of ABM (presented in 2.2) as well as, the different classes of ML algorithms and the respective tasks performed with these (presented in 2.3). The resulting classification scheme is shown in Table 1. Note that the order or proximity of characteristics is not intended to imply predestined theoretical linkages between dimensions. These linkages are to be determined through our analysis of the applied research in our sample. We rely on the expertise of our team of authors to classify each publication along the four dimensions, starting with the abstracts and confirming it by going through the full papers. Again, this step is double-checked. Of course, a publication is not always one thing or the other and we find a number of publications highlighting more than one application of differing ML algorithms to enhance the ABM. In case a research paper cannot be assigned clearly to only one element in a dimension of our classification scheme, we decide to co-classify articles for each use case of ML in ABM. For example, Nogueira and Oliveira (2007) employ an RL algorithm to model independent agents but also equip agents with the ability to perceive the environment using a clustering algorithm. Thus, we classify the paper for each of the two different use cases of ML. Similarly, one single ML algorithm may result in more than one use case by simultaneously improving computational efficiency and helping to increase the validity of the model, as is the case in Takadama, Kawai, and Koyama (2008). Again, we double-count this entry for the two different goals. We acknowledge that the boundaries between the steps of the ABM process may be floating (or circular) and that the pursued goals of using ML surely are interconnected. However, we classify the articles to the best of our understanding by focusing on the major objectives stated by the authors of the respective publications.

Table 1: Classification Scheme Used to Classify the Relevant Literature and the Most Important Use Cases.

| **ABM steps 1-4** | **Overall Goals I-IV** | **Applied Methods a-d** | **MLtasks R,CA,P,CU,AD,AR,D,O** |
|---|---|---|---|
| (Which are the relevant steps typically performed in ABM?) | (What are the overall goals ML is so far used in and for ABM?) | (Which ML methods are so far applied to reach the overall goal?) | (What is the major task of the applied algorithm?) |
| **1. Empirical Parametrization**<br>(i.e. chose target, make assumptions based on empirical data) | **I. improve accuracy**<br>i.e. the number of data points and (quality of) assumptions built into (or inform) the model) | **a) supervised**<br>(e.g. regression or classification) | **R - Regression:** (e.g. predicting a continuous quantity)<br>**CA - Classification:** (e.g. predicting a discrete class label)<br>**P - Time-series forecasting:** (e.g. to predict future observations) |
| **2. Construction / Implementation of Computational Model**<br>(i.e. writing a special computer program to implement model design in silico) | **II. improve computational efficiency**<br>(i.e. reducing computational time / load, reduce comp. time through ML surrogates, reduce comp. load by excluding less relevant parameters) | **b) unsupervised**<br>(e.g. clustering, time series analysis or anomaly detection) | **CU - Clustering:** (e.g. clustering data points in groups)<br>**AD - Anomaly Detection:** (e.g. identifying unexpected items or events in data sets)<br>**AR - Association Rules:** (e.g. discovering interesting relations between variables in large databases)<br>**D - Dimensionality Reduction** (e.g. reducing the number of random variables under consideration) |
| **3. Validation & Verification**<br>(i.e. check whether the simulation is a good model of the target and checking that a program does what it was planned to do a sensitivity analysis of model behavior (w.r.t. assumptions made to test robustness of model) | **III. improve validity of the model**<br>(i.e. increasing the validity of a simulation model (check whether the simulation is a good model of the target)) | **c) semi-supervised**<br>(e.g. time series analysis) | |
| **4. Output Analysis**<br>(i.e. finding emerging patterns for a combination of relevant parameters to derive a narrative/implication from output data) | **IV. improve understanding**<br>(i.e. understanding of model behaviors, analysis of outcome data to find important parameters, clusters, conditions, causalities) | **d) reinforcement learning**<br>(e.g. optimization) | **O - Optimisation:** (Maximisation) –> Reinforcement Learning |

# 4 Results and Discussion

## 4.1 Main Results

In the following section, we're first going to briefly present some descriptive statistics as the number of publications over time, the most important journals researchers published in, as well as the most prominent

keywords. Second, we're going to present the two most important use cases by giving some examples from our sampled literature.

Figure 1: Number of Publications over Time.

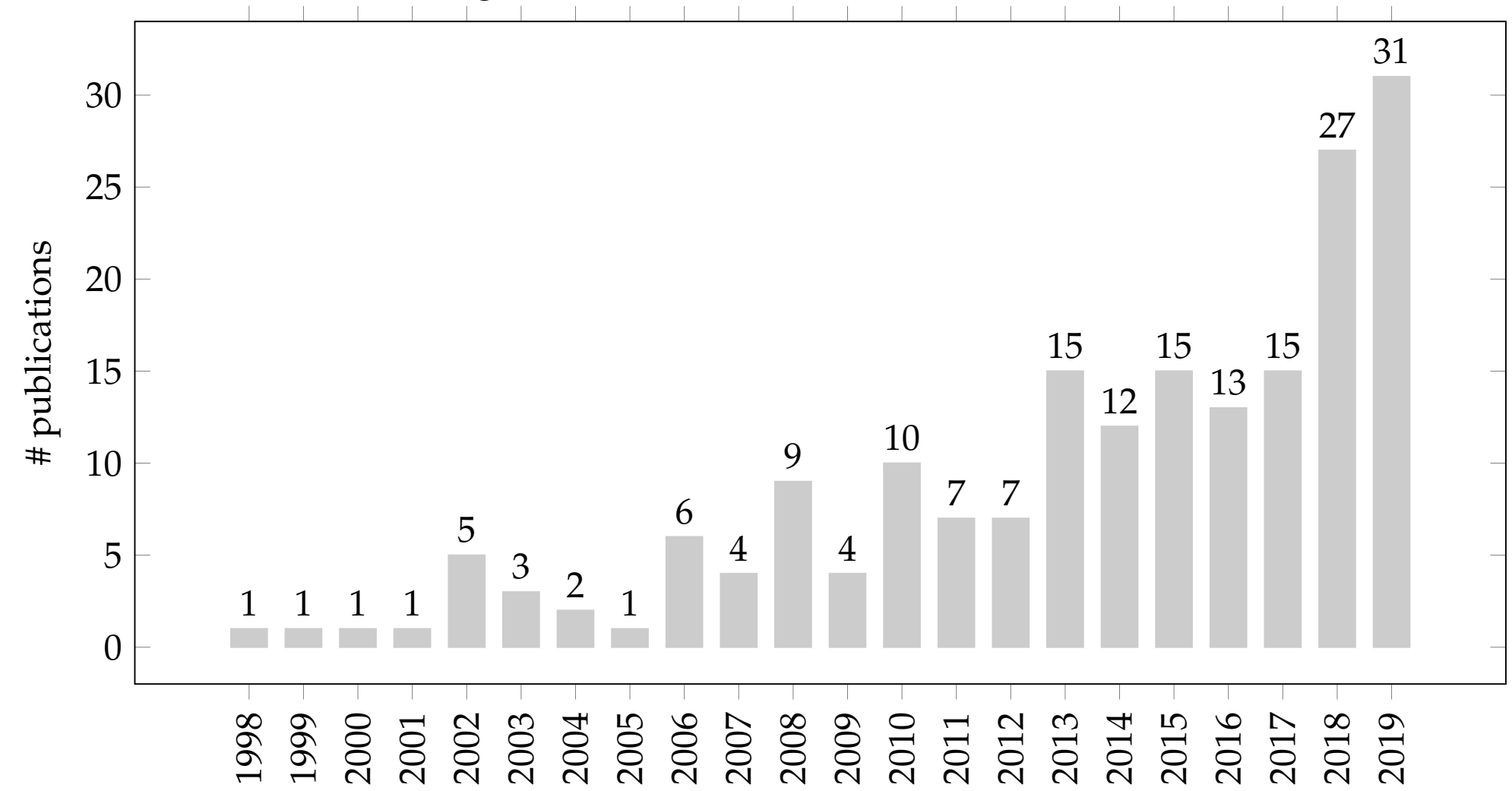

Figure 1 shows how the number of papers dealing with the application of ML in and for ABM has increased over time. More than half of all papers in this topic have been published within the last five years. This increase over time shows the growing interest in and importance of this topic. Not only has the number of publications increased in general, but we can also see a strong research interest of many different disciplines.

Figure 2: Different Journals Papers are Published in.

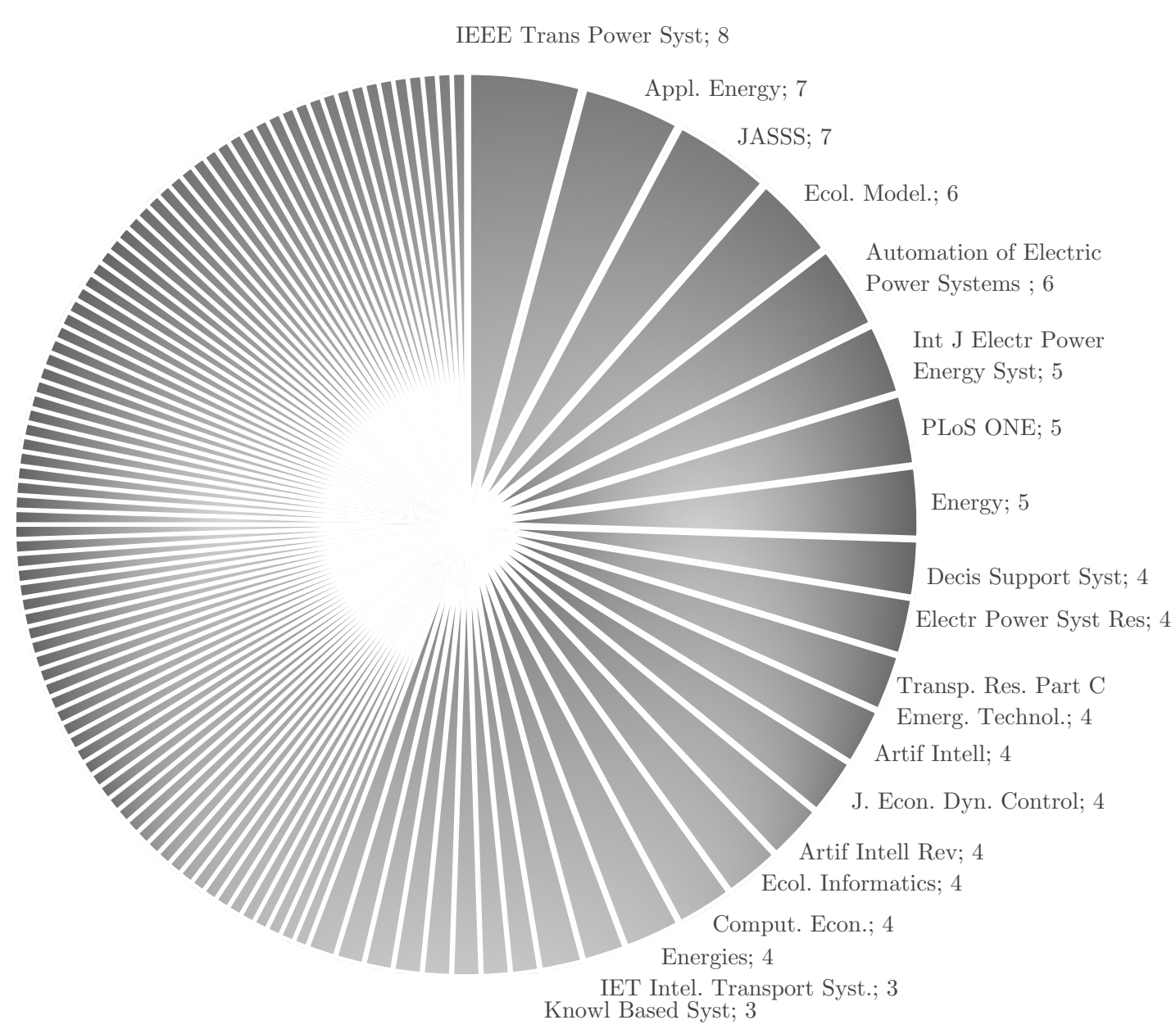

Looking at the different journals presented in Figure 2 shows that there are no dominant journals, but we find 190 relevant papers published in 112 different journals. Most articles are published in IEEE Transactions on Power Systems (8), Applied Energy (7), Journal of Artificial Societies and Social Simulation (7), Ecological Modelling (6), Dianli Xitong Zidonghue / Automation of Electric Power Systems (6),

International Journal of Electrical Power and Energy Systems (5), PLoS ONE (5), Energy (5), Decision Support Systems (4), Electric Power Systems Research (4), Transportation Research Part C (4), Journal of Artificial Intelligence (4), Journal of Economic Dynamics and Control (4), Artificial Intelligence Review (4), Ecological Informatics (4), Computational Economics (4), Energies (4), IET Intelligent Transport Systems (3), and Knowledge-Based Systems (3). Despite having publications in many different journals, we find a high proportion of applications of ML in and for ABM in research related to the energy system. This high number of different journals without doubt shows the heterogeneity of possible applications and disciplines and the enormous potential ML might play in and for many different kinds of ABMs.

To get a more detailed understanding and to address our main research question, in Figure 3, we classify, the relevant literature according to the scheme presented in Table 1. By looking at our classification scheme we directly see that most applications address the ABM step 2) construction of the mathematical model (155). Concerning the most important overall goal, a similarly clear picture emerges. Most studies by far use ABM in order to improve the accuracy of their model (189) (e.g. by improving the quality of assumptions build into their model). Another goal is to apply ML to improve the modelers understanding of the model output and the input-output-relationships (27). Less important seems to be the improvement of the computational efficiency (9), or the validation and verification of the model (6). The ML method applied most is reinforcement learning (137), followed by supervised learning (48), unsupervised learning (11), or a mixture between the two (1). Regarding the major task of the applied algorithm, by far the most applications use ML for optimisation tasks (144). This is followed by classification tasks (32), regression (17), clustering (7), time-series forecasting (2), as well as only one application of ML in ABM for anomaly detection, association rules and dimensionality reduction, respectively. This first classification of use-cases found in the literature shows that most papers deal with the application of ML in and for ABM to improve the implementation of the model itself, e.g. the kind of agents and how they are implemented in the model, how they learn in the model, and to improve the accuracy of the model, e.g., by increasing the number of data points and (quality of) assumptions built into the model. Most of the time, authors use reinforcement learning to support and use ML to perform optimisation tasks. By looking at the papers in more detail, we find that ML in ABM is mainly used for two broad cases: First, the modelling of adaptive agents equipped with experience learning and, second, the analysis of outcomes produced by a given ABM. However, looking at the number of use cases identified as depicted in Figure 3, the predominant application seems to be the former: 137 cases deal with the use of reinforcement learning in modelling agents versus 27 cases related to the output analysis. As already mentioned above, the application of RL in ABM is strongly represented in MAS-based literature stemming from energy systems research but can be found across all subject areas. Already by looking at the most important keywords, we see that some of the most frequent keywords are reinforcement learning and Q-learning, showing this strong dominance of modelling adaptive agents equipped with experience learning. To get a better understanding, in the following, we're going to present these two use cases by giving some examples from our sampled literature.

Figure 3: Different Applications Found in the Literature.

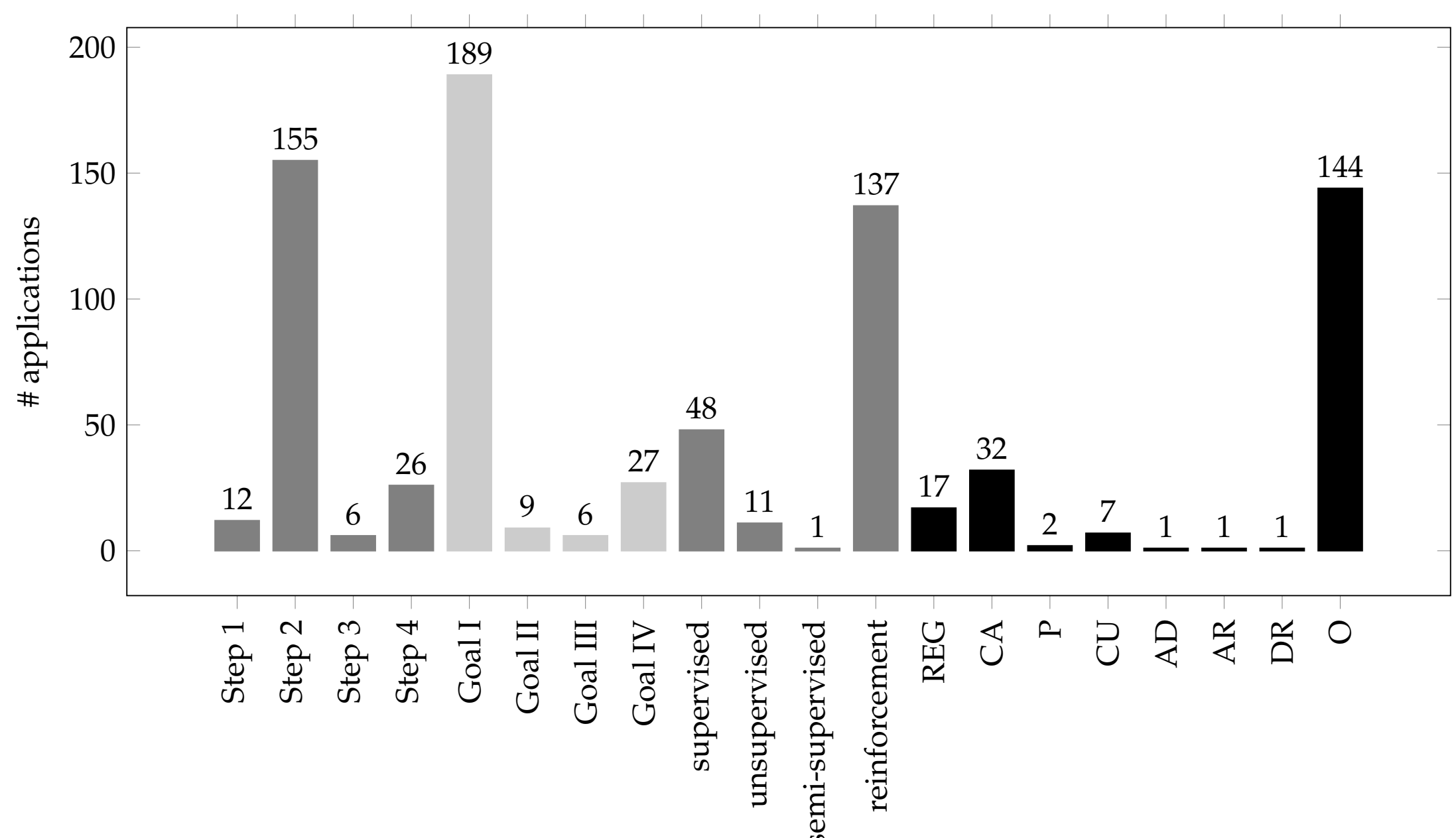

**On the Need for Reinforcement Learning to Create Adaptive Agents**

We find that the most prevalent use of ML concerns the issue of accuracy addressed through behavioral modelling and utilizes RL algorithms as micro-surrogates to enhance behavioral capabilities of agents. By using multi-agent reinforcement learning (MARL), the accuracy of models may be increased by allowing agents to experimentally learn through assessing the outcomes of their actions. On the need for these RL algorithms in such simulations, Busoniu, Babuska, and De Schutter (2008, p.156) contend the following:

> *"Although the agents in a multi-agent system can be programmed with behaviors designed in advance, it is often necessary that they learn new behaviors online, such that the performance of the agent or of the whole multi-agent system gradually improves (...) This is usually because the complexity of the environment makes the a priori design of a good agent behavior difficult or even impossible. Moreover, in an environment that changes over time, a hardwired behavior may become unappropriate."*

The most common type of algorithm applied for this purpose are Q-learning algorithms. Q-learning agents are usually modelled through a Q-value matrix, a probability distribution matrix, a reward function, an action set, and a state set. In each tick, an agent receives a state and reward from the environment as factors to update the Q-value and probability distribution matrices before performing a new action and producing an output exerted into the environment (Yin, Yu, & Zhou, 2018). Several studies in our sample claim that this may closely mimic human-like behavior and excels in comparison with theoretically modelled agents (Takadama, Kawai, & Koyama, 2008; Tang, 2008). By analyzing our sample, we follow Busoniu, Babuska, and De Schutter (2008) in distinguishing two main approaches: that is, the single-agent and the multi-agent case of equipping agents with RL capabilities. In our sample, the former is the more common case. Here,

RL is used to model reinforced learning of individual agents adjusting their actions to optimize individual goals (Vista et al., 2014), for example by adjusting prices (Wang, Wu, & Che, 2019; Yanagita & Onozaki, 2008), risk aversion levels (Esmaeili Aliabadi, Kaya, & Sahin, 2017) or improving forecasting abilities (Li & Shi, 2012). Second, for the multi-agent case, we see (cooperative) reinforcement learning applied to collectively optimize outcomes on the macro level. In this case, the transition of states and resulting rewards in the Q-function are dependent on the joint action of all (or groups of) agents (Busoniu, Babuska, & De Schutter, 2008). Cooperation may be achieved in different ways. Kofinas, Dounis, and Vouros (2018) show that independent Q-learning agents can collectively coordinate by sharing information on their own states with other agents instead of only producing outputs (see also Yu et al. (2016)). Using a slightly different approach in a study on traffic signal controls, Zhu et al. (2015) introduce a junction-tree-algorithm (JTA) designed for coordinated reinforcement learning in order to determine the best joint actions of agents. Depending on the overall goal of the researcher and the underlying assumptions, such learning agents can indeed be a complement to or improvement of the ABM. Even a combination of the single-agent and multi-agent case may enable a more realistic approximation of reality. Challenges with implementing MARL into a simulation model may be caused by the curse of dimensionality of the Q-learning algorithm. Calculating all possible combinations of an increasing number of state-action pairs and/or an increasing number of agents may cause exponential growth of needed computational resources. Moreover, coordinating actions when facing dynamic learning problems dependent on other agents' actions can prove a difficult task in model construction, i.e. facing the problem of balancing exploration and exploitation behavior of single agents (Busoniu, Babuska, & De Schutter, 2008).

**On the Need for Analyzing Model Outputs**

While an absolute number of 27 articles seems negligible, the second most prominent use case of ML in ABM found within our sampled literature concerns the goal of output analysis (almost 15% of articles in our sample). Taking the importance of output analyses into account, this rather small number of applications seems somewhat surprising. As Edali and Yücel (2019, p. 1) frame it:

> *"The main motivations behind using agent-based modeling are mainly to understand the dynamics of the system and to observe how the system behaves in the presence of interventions (e.g., new policies, parameter changes) (...). In this respect, discovering the relationship between inputs and outputs of agent-based models is the ultimate way of providing insights into understanding the dynamics of the system being modeled."*

In the same vein, Pereda, Santos, and Galán (2017, p. 3) argue that

> *"[...] the task of analysing the relation between the model output and parameters is usually not simple or easy. The difficulty is greater as the amount of parameters is larger, complicating the use of the traditional graphical techniques to draw inferences. In general, but particularly in models with high dimensional parameter spaces, machine learning techniques* *can be usefully applied to analyse ABM models."*

Despite the rather small number of cases, we found such applications of ML to analyse ABM models. We discover Perry and O'Sullivan (2018) making a compelling and representative case for the use of machine learning to derive interpretable narratives out of models that generate overwhelmingly large outputs. That is, by providing "powerful ways to identify narratives from individual trajectories and to isolate the circumstances under which they might arise" (Perry & O'Sullivan, 2018, p. 814). In a model on hunter-gatherer foraging, they apply a two-stage approach for output analysis of ABM. First, a clustering process is conducted to find groups of similar scenarios in the output data produced by a model. Perry and O'Sullivan (2018) use dynamic time warping (DTW) and metric multidimensional scaling in conjunction with a k-means clustering algorithm. Second, the identified clusters in the output data need to be connected to specific parameterizations increasing the likelihood of producing specific scenario clusters. To this end, Perry and O'Sullivan (2018) perform a classification procedure supported by decision-tree-based random forest algorithms and visualize their finding by using partial dependence plots (PDP). PDP allow to graph the marginal effects of up to two varied variables (ceteris paribus) on the predicted outcome of the ML algorithm (Zhao & Hastie, 2019). Li et al. (2013) prefer a DBSCAN (density-based spatial clustering of applications with noise) clustering algorithm over k-means for analyzing a model on geographic economic development, pointing towards the benefits of not being required to determine the number of clusters in advance and that the algorithm is generally performing well for spatial data. This two-step process is representative of other studies in our sample. For example, for analyzing predator-prey dynamics (Khater, Murariu, & Gras, 2014; Scott, MacPherson, & Gras, 2018). Some studies do not carry out a ML clustering step but proceed to employ random forest (i.e. decision trees) just the same to explain statistical variations in model outcomes (Ma et al., 2019; Vahdati et al., 2019). Yet again, other studies seek to simply identify the most important parameters that are driving one specific model outcome, e.g., by employing naive Bayes and perceptron classifiers (Kamar, Gal, & Grosz, 2013). Strien et al. (2019) use a more antiquated type of ML classification algorithm in the form of support vector machines to quantitatively distinguish equilibria of positive or negative growth in the model results. However, they concede that the classification error amounted to around 18%.

### 4.2 Discussion

Investigating the papers on the application of ML in and for ABM in general and the two most prominent use cases in particular we see that (i) the extent to which ML is used in and for ABM increased tremendously over the last years, and (ii) indeed, it actually seems to be possible to address current challenges of ABM by means of ML – and this in many diverse ways. To explicitly address our second research question, in this sub-chapter we will briefly summarize and discuss the potential strengths and weaknesses of using ML in and for ABMs.

The major strengths we found in the literature are the following four: 1) ML may complement ABMs by increasing the explanatory power of ABMs, 2) improving the computational efficiency, 3) by improving validation and verification of ABMs, and 4) by enhancing the analyst's understanding of input-output-relationships and of factors influencing model outcomes. While acknowledging these strengths, we must not forget that the application of ML in and for ABM come at certain costs and different aspects or problems have to be kept in mind. Indeed, ML techniques, as for instance reinforcement learning, enable ABMs to become a more accurate representation of the real world, thereby increasing explanatory power. That is, by providing tools to analyze models that are larger in scale and scope, by informing model building through parsing real-world data and by increasing the adaptivity of agents through experiential learning. This directly addresses the behavioral modelling challenge (e.g. by modelling learning agents by means of reinforcement learning) as well as the large-scale agent-based modelling challenge (e.g. creating macro-surrogates as Edali and Yücel (2019) or Lamperti, Roventini, and Sani (2018)). However, more accurate representations of reality might come at the cost of losing focus and the ability to understand the model, e.g. by creating intelligent, yet black-box ABMs. What is more, optimizing reinforcement learning agents sometimes might contradict the notion of agents following simple rules of thumb and again trap in the mainstream neoclassical assumption of rational optimizing agents.

The possibility of improving the computational efficiency and reducing computational time of ABMs leads to bright excitement among modelers. Increasing the computational load of models, e.g., by excluding less relevant parameters identified through ML algorithms, or substituting (parts of) the ABM with faster ML surrogates emulating the model, seem to be quite promising in addressing the large-scale agent-based modelling challenge. However, there is no free lunch: Using ML may be complicated and time consuming. Choosing suitable algorithms and settings is not trivial and not all algorithms are equally applicable for all cases. Surrogates require training (e.g. with data generated by ABMs) as well as cross-validation, testing, etc. What is more, excluding less relevant parameters may increase efficiency but for the sake of losing detail. It has to be kept in mind, that we might face a trade-off between computational efficiency and accuracy.

ML techniques can support the modeler by enhancing the understanding of input-output- relationships and of the factors influencing model outcomes, e.g. by clustering outcomes or by measuring the importance of classified parameters. This directly addresses the simulation analytics challenge as well as the research challenge of the large-scale agent-based modelling challenge. However, it has to be kept in mind that when creating models in which we are no longer able to understand the processes involved, we cannot understand these artificial complex systems any better than we understand the real ones (Axtell & Epstein, 1994; Gilbert & Terna, 2000). At this point, even if in the end, we get some interesting input-output- relationships, if we do not understand the dynamics behind these relationships the model decays to a meaningless construct without any real scientific value, which in the best case can be used to visualize our little understanding of the matter. In addition, when relying too much on the 'most important' parameters influencing model outcome, we run into danger of losing explanatory power by aggregating to high level descriptors. This might lead to a situation in which the original advantages of ABMs, are overshadowed by the disadvantages of applying ML.

Hence, despite the many benefits of applying ML in ABM, it is important keep in mind some general obstacles that we may encounter when using ML for ABM. First, we have to keep in mind that the field of machine learning covers many different approaches, techniques and algorithms. Finding an adequate algorithm requires a rather specific set of experience and still may include a time-consuming trial and error process where we test different algorithms and settings for each algorithm. Second, even after identifying an appropriate algorithm, applying it may still be non-trivial and time-consuming task. For instance, the training time to train just one neural network can range between few minutes and several hours or more and consequently requires considerable computational capacities. Third, applying ML in ABM for a specific purpose often includes the risk of creating unintended side effects that cause new problems. For example, when using micro surrogate models to increase the accuracy of the agents' behaviour, we have to consider that the computational time needed to execute the model increases considerably and understanding the model behaviour becomes more complex. In a similar vein, if our aim is to increase the computational efficiency of our ABM for example through using macro-surrogates, we may lose important details of the model results and thus explanatory power of the whole ABM model.

## 5 Summary and Conclusion

The interest in the method of agent-based modelling has grown tremendously in recent years, being increasingly applied by different researchers from various disciplines in many diverse fields. However, especially in larger and more complex models, modelers often face problems related to the increase in used and produced (big) data. Within the last few years, an ever-increasing number of interesting publications propagated the seemingly endless possibilities of addressing these problems by using machine learning techniques for and in agent-based simulation models. To get a more comprehensive overview over how these techniques actually are applied and in which contexts, as well as to assess possible gains and pains in applying ML in and for ABM. In this paper we conduct a Systematic Literature Review of the most important contributions in this context. In our analysis, we find that ML is an open field showing fast development and offering many future possibilities. This being the case, ML offers a diversity of promising tools to support and complement ABMs in many different ways. In more detail, we find that ML is mainly used in ABM for two broad cases: First, the modelling of adaptive agents equipped with experience learning and, second, the analysis of outcomes produced by a given ABM. While these are the most frequent applications, there also exist many more interesting applications. By shedding light on the possible pains and gains of the combination of these two promising methods, it has to be kept in mind that the actual research goal / objective of employing ML in and for your ABM (motivated by efficiency or analytical depth) is decisive. We conclude that ML may complement and support ABM in cases in which the aim is to increase tractability of large-scale macro models, in cases in which the overall goal is handling models with high dimensionality, multi-modal outputs but less important boundary conditions, and especially in cases in which we seek to model adaptive agents (i.e. to find optimal strategies), ML can be used to determine agents' behavior (goal-oriented agents).

The transferability of our results is limited for several reasons. The most important one is that we limit our research fields to only a few subject areas, there might be a bias, e.g., towards energy system research related applications. Therefore, future research should both broaden the analysis by incorporating even more subject areas, but also conduct more in-depth analyses, e.g., of different use cases and their applicability to other problems. Nonetheless, as previous studies only presented a present a fragmented view on some potential uses, this study clearly advances research by giving a more comprehensive overview.

# Appendix A

Table 2: Search Query.

| | |
|---|---|
| **Search query** | TITLE-ABS-KEY ( ( "machine learning" OR "machine intelligence" OR "machine translation" OR "supervised learn*" OR "unsupervised learn*" OR "reinforcement learn*" OR "decision tree" OR "random forests" OR "naive Bayes" OR "support vector machine*" OR "k-means" OR "k-nearest neighbor" OR "low-pass filter" OR "apriori alg*" OR "gradient descent" OR "Q-learning" OR "neural net*" OR "convolutional" OR "recurrent" OR "backprop*" OR "feedforward" OR "deep learn*" OR "self-organizing map" OR "self-organising map" OR "restricted Boltzmann machine" OR "deep belief net*" OR "generative adversarial net*" OR "data-min*" OR "natural language processing" ) AND ( "agent-based model*" OR "agent based model*" OR "multi-agent system*" OR "multi agent system*" OR "multi-agent sim*" OR "multi agent sim*" OR "agent-based sim*" OR "agent based sim*" OR "multi-agent-based sim*" OR "multi agent based sim*" OR "agent-based social sim*" OR "agent based social sim*" OR "individual-based model*" OR "individual based model*" OR "individual-based configuration model*" OR "individual based configuration model*" ) ) AND ( LIMIT-TO ( DOCTYPE , "ar" ) ) AND ( LIMIT-TO ( SUBJAREA , "SOCI" ) OR LIMIT-TO ( SUBJAREA , "BUSI" ) OR LIMIT-TO ( SUBJAREA , "ENVI" ) OR LIMIT-TO ( SUBJAREA , "ENER" ) OR LIMIT-TO ( SUBJAREA , "AGRI" ) OR LIMIT-TO ( SUBJAREA , "ECON" ) ) |
| **Found articles** | 397 |
| **Excluded articles** | 207 |
| **Final sample** | 190 |